\documentclass[conference]{IEEEtran}
\IEEEoverridecommandlockouts
\usepackage{cite}
\usepackage{amsmath,amssymb,amsfonts}
\usepackage{algorithmic}
\usepackage{graphicx}
\usepackage{textcomp}
\usepackage{xcolor}
\def\BibTeX{{\rm B\kern-.05em{\sc i\kern-.025em b}\kern-.08em
    T\kern-.1667em\lower.7ex\hbox{E}\kern-.125emX}}
    
\usepackage{pgfpages}
\pgfpagesuselayout{resize to}[a4paper]   
    
\begin{document}

\title{Data Synopses Management based on a Deep Learning Model
}

\author{\IEEEauthorblockN{Panagiotis Fountas}
\IEEEauthorblockA{\textit{Department of Informatics and Telecommunications} \\
\textit{University of Thessaly}\\
Papasiopoulou 2-4, 35131 Lamia Greece \\
pfountas@uth.gr}\\
\IEEEauthorblockN{Christos Anagnostopoulos}
\IEEEauthorblockA{\textit{School of Computing Science} \\
\textit{University of Glasgow}\\
Lilybank Gardens 18, G12 8RZ Glasgow UK \\
christos.anagnostopoulos@glasgow.ac.uk}\\
\and
\IEEEauthorblockN{Kostas Kolomvatsos}
\IEEEauthorblockA{\textit{Department of Informatics and Telecommunications} \\
\textit{University of Thessaly}\\
Papasiopoulou 2-4, 35131 Lamia Greece \\
kostasks@uth.gr}\\


}

\maketitle

\begin{abstract}
Pervasive computing involves the placement of processing units and 
services close to end users to support intelligent applications that will facilitate their activities.
With the advent of the Internet of Things (IoT) and the Edge Computing (EC),
one can find room for placing services at various points in the interconnection of the aforementioned infrastructures. 
Of significant importance is the processing of the collected data to provide 
analytics and knowledge. 
Such a processing can be realized upon the EC nodes that exhibit increased computational capabilities compared to IoT devices.
An ecosystem of intelligent nodes is created at the EC giving the opportunity to 
support cooperative models towards the provision of the desired 
analytics.
Nodes become the hosts of geo-distributed datasets formulated by the IoT devices reports.
Upon the datasets, a number of queries/tasks can be executed either locally or remotely.
Queries/tasks can be offloaded for performance reasons to deliver the most appropriate response. 
However, an offloading action should be carefully designed being always aligned with the 
data present to the hosting node.
In this paper, we present a model to support the cooperative aspect in the EC infrastructure. 
We argue on the delivery of data synopses distributed in the ecosystem of EC nodes making them capable to take offloading decisions fully aligned with data present at every peer.
Nodes exchange their data synopses to inform their peers.
We propose a scheme that detects the appropriate time to distribute the calculated synopsis
trying to avoid the network overloading especially when synopses are frequently extracted due to 
the high rates at which IoT devices report data to EC nodes.
Our approach involves a Deep Learning model for learning the distribution
of calculated synopses and estimate future trends.
Upon these trends, we are able to find the appropriate time to deliver synopses to peer nodes.
We provide the description of the proposed mechanism and evaluate it 
based on real datasets.
An extensive experimentation upon various scenarios reveals the pros and cons of the approach
by giving numerical results.
\end{abstract}


\section{Introduction}
Pervasive computing targets to the creation of smart environments around end users
saturated with computing and communication capabilities to support novel applications.
Pervasive services aim to be invisible, however, intelligent enough to facilitate users 
activities.
Today, we are witnessing the provision of huge infrastructures where 
pervasive applications can be hosted.
Such infrastructures deal with the Internet of Things (IoT) and Edge Computing (EC).
Both of them try to `surround' end users with smart devices, collect and process data to create knowledge adopted by various applications.
It becomes obvious that in this new era of the Web, there are numerous opportunities 
to support intelligent and invisible services 
in a close distance with users. Hence, we are able to 
reduce the latency in the provision of the discussed services and increase the performance.
The first `actor' in this setting is the IoT device that may directly interact with 
users and their environment to 
collect data and perform simple processing activities 
\cite{najam}.
IoT devices can, then, report their data in an upwards mode, to the 
EC infrastructure and Cloud for further processing.
EC involves an ecosystem of heterogeneous nodes that become the hosts of the collected data and act as processing points to deliver analytics and knowledge
\cite{najam}.
We can observe a high number of distributed datasets present at the network edge opening the room 
for defining advanced services and support real time applications. 
The aim is to serve users or applications requests in the minimum time with the maximum performance.
As the maximum performance we denote the provision of responses that fully match to the incoming requests.
Obviously, responses are provided upon the available data and should be aligned with them.

As the EC supports a distributed environment with numerous datasets present at the ecosystem of nodes, the use of cooperative models is imposed to make nodes capable of exchanging data, queries/tasks, etc. 
The interaction between EC nodes aims at detecting the appropriate line of actions to efficiently respond to the incoming requests for processing.
The reaserch community has already focused not only on the management of 
queries/tasks 
\cite{karanika}, \cite{kolomvatsos1}, \cite{kolomvatsos2}, \cite{kolomvatsos3}, \cite{kolomvatsos4} 
but also on the management of the collected data \cite{amrutha}.
However, due to the distributed nature of the EC, nodes should have a view on the 
data present in peers especially when we want to support efficient decision making locally.
For instance, a data allocation action demands for the knowledge of the remote data at least in the form of synopses.
Data synopses can be exchanged between EC nodes without burdening the network
as they usually convey high level statistical information about the 
available data.
The delivery of synopses seems to be more efficient than the exchange of large pieces
of data, i.e., data migration between nodes \cite{bellavista}.
Actually, we have two solutions for responding to queries/tasks requests when 
the relevant data are note present at the node receiving the request. 
The first solution deals with the queries/tasks migration upon the decision of finding the most appropriate node as seen by the corresponding synopsis.
The second solution deals with the migration of the relevant data from the owner/peer to 
the node receiving the request.
Evidently, the former model should be supported by an intelligent mechanism
for exchanging the necessary statistical information for the data present in the ecosystem 
while the latter scheme burdens the network with large messages 
increasing the possibility of bottlenecks. 

In this paper, we support the autonomous nature and the cooperation between 
EC nodes to serve queries/tasks demanded by users or applications.
We focus on the first of the aforementioned solutions 
(i.e., queries/tasks migration) and propose a scheme for exchanging 
data synopses in the ecosystem to efficiently support decisions related to 
offloading actions.
Synopses are updated every time new data arrive 
in an EC node, however, they should distributed 
when their `magnitude' exhibit that significant new information is present.
We propose a monitoring mechanism for the updated synopses
and a model that detects when significant changes are present at every dataset.
When this is true, we decide to deliver the synopses to peer nodes to have them 
informed about the new status of every dataset.
We rely on a Deep Machine Learning (DML) model 
to learn the distribution behind the calculated synopses
being able to estimate their future trends.
Hence, in a proactive manner, we are able to estimate the
appropriate time for delivering the updated synopses.
More specifically, we adopt a Long Short Term Memory (LSTM) network which is a specific type of Recurrent Neural Networks (RNNs) \cite{goodfellow}.
The adopted LSTM is
capable of 
incorporating 
data from the previous step to the 
upcoming steps of processing.
Hence, LSTMs are capable of identifying dependencies on data 
that `legacy' neural networks cannot do.
The detection of such dependencies are critical in our scenario as 
synopses are updated in an `incremental' manner, i.e., 
new data arrivals are affecting the statistical information of datasets that is 
related to the previously delivered synopses. 
We consider the trade off between the frequency of synopses distribution and the 
`magnitude' of updates. 
We can accept the limited freshness of updates for gaining
benefits in the performance of the network.
We also define an uncertainty driven model under the principles 
of Fuzzy Logic (FL) \cite{fuzzy} to decide when an EC node should 
distribute the synopsis of its dataset.
The uncertainty is related to the `threshold' (upon the differences 
of the available data after getting reports from IoT devices) over which 
the node should disseminate the current synopsis.
We monitor the `statistical significance'
of synopses updates before we decide to distributed them in the network.
We consider the trade off between the frequency of the distribution and the 
`magnitude' of updates. We can accept the limited freshness of updates for gaining
benefits in the performance of the network.
We apply our scheme upon past, historical observations (i.e., synopses updates)
as well as upon future estimations.
Both, the view on the past and the view on the future (derived by the proposed LSTM) are fed into our Type-2 FL System (T2FLS)
to retrieve the \textit{Degree of Distribution} (DoD).
Two DoD values (upon historical values and future estimations) are smoothly aggregated through a 
geometrical mean function \cite{mesiar} to finally decide the dissemination action.
Our contributions are summarized by the following list:
\begin{itemize}
	\item We support monitoring activities for detecting the magnitude of the updated synopses;
	\item We deliver an LSTM for learning the distribution and dependencies on continuous updates of data synopses for estimating their future realizations; 
	\item We propose an uncertainty driven model for detecting the appropriate time to distribute data synopses to peers;
	\item We report on the experimental evaluation of the proposed models through a large set of simulations.
\end{itemize}

The paper is organized as follows.
Section \ref{related} presents the related work while Section 
\ref{preliminaries} formulates our problem and provides the 
main notations adopted in our model.
In Section \ref{uncertainty}, we present the envisioned mechanism and 
explain its functionalities.  
In Section \ref{evaluation}, we deliver our experimental evaluation
and conclude the paper in Section \ref{conclusions} by presenting 
our future research directions.

\section{Related Work}
\label{related}
Resource management at EC has been studied in the past to reveal 
the requirements for hosting and processing data.
This is because data processing demands for specific 
resources according to the complexity of the requested
queries/tasks. The appropriate allocation of the available resources
will guarantee the increased performance 
and the timely provision of the outcomes.
A number of efforts try to 
deal with the resource management problem
\cite{anglano}, \cite{Cherrueau}, \cite{Shekhar}, \cite{wang}.
Their aim is to address the challenges on how we can 
offload various tasks/queries and data to EC nodes taking into consideration 
a set of constraints, e.g., time requirements, communication 
needs, nodes' performance, the quality of the provided responses and so on
and so forth. 
A relevant study in the domain reveals 
that processing nodes may adopt the following three (3) schemes to 
perform the execution of queries/tasks \cite{wang}:
\begin{itemize}
	\item An integration model for aggregating data reported by multiple devices \cite{ref17}.
EC nodes have the opportunity to locally process the collected data before they transfer them to the Cloud. This approach limits the time for the provision of the final response as the processing is performed in close distance with end users; 
	\item A `cooperative' scheme though which EC nodes interact with other devices (e.g., IoT devices, EC nodes) having processing capabilities to offload a subset of tasks \cite{ref21}. In any case, the distance between the interacting devices should be low, otherwise,
their interaction may be problematic;
	\item A `centralized' approach where EC nodes act as execution points for queries/tasks offloaded by IoT devices \cite{ref26}. This approach sees EC nodes having increased computational resources compared to IoT devices, thus, they can perform more complicated processing. Arguably, EC nodes should incorporated a monitoring mechanism to detect possible overloading cases and take specific mitigation actions.
\end{itemize}
Additionally, for speeding up the processing at the EC nodes while being aligned with the requirements of requests, various efforts have proposed 
the use of caching \cite{ref32}, 
context-aware web browsing \cite{ref33} 
and pre-processing actions \cite{ref34}.

Evidently, queries/tasks are executed upon huge volumes of data.
The processing of large scale data demands for efficient techniques 
to timely deliver the outcomes.
The support of synopses management is already identified by the research community as a means
for having a view upon the data avoiding to perform time consuming activities,
Synopses convey statistical information about the underlying data
\cite{aggarwal} and can be maintained in an incremental
approach.
The research community has connected the term `synopsis' with 
(i) approximate query estimation \cite{Chakrabarti}: 
the target is to estimate, in real time, responses given the query and without having any view on data. 
Obviously, the final goal is to detect the data that  
better `match' to the incoming queries;
(ii) approximate join estimation \cite{alon}, \cite{dobra}: join operations are usually time concuming and more complex compared to other types of operations (e.g., a simple select 
upon the available data). Hence, approximate solutions may limit the time required to conclude any join operation taking into consideration the trade off between the accuracy of results and the 
conclusion time;
(iii) aggregates calculation \cite{Charikar}, \cite{Cormode}, \cite{Gehrke}, \cite{Manku}: the aim is to provide aggregate statistics over the available data;
(iv) data mining schemes \cite{Aggarwal1}, \cite{Aggarwal2}, \cite{Schweller}: 
there are services demanding for synopses instead of the individual data points, e.g., clustering, classification. This means that any decision is retrieved upon the high level statistical information for the available data.
In any case, the adoption of data synopses aims at the processing of 
only a subset of the actual data.
Synopses act as `representatives' of data and usually involve 
summarizations or the selection of a specific subset
\cite{Lakshmi}.
Any limited representation may heavily reduce the need 
for increased bandwidth of the network and 
can be easily transferred in the minimum possible time.
Examples techniques for the delivery of synopses deal with
sampling \cite{Lakshmi},
load shedding 
\cite{Babcock1}, \cite{Tatbul},
sketching \cite{Babcock}, \cite{Muthukrishnan} and 
micro cluster based summarization \cite{Aggarwal1}.
The easiest one is sampling. It targets to the probabilistic selection of a subset of the actual data.
Obviously, the appropriate selection of samples plays a significant role in the 
success of the decision making model where samples are processed. 
Load shedding aims to drop some data when the system identifies a high load, thus, to 
avoid bottlenecks.
Sketching involves the random projection of a subset of features that describe the data 
incorporating mechanisms for the 
vertical sampling of the stream.
Micro clustering targets to the management of the multi-dimensional aspect of any data stream 
towards to the processing of the 
data evolution over time.
Other statistical techniques 
are histograms and wavelets \cite{aggarwal}.

\section{Preliminaries and Problem Description}
\label{preliminaries}
We focus on the ecosystem of EC nodes that exhibit cooperative behaviour towards the execution of the 
received queries/tasks. 
Without loss of generality, we assume $N$ EC nodes depicted by the 
set 
$\mathcal{N} = \left\lbrace n_{1}, n_{2}, \ldots, n_{N} \right\rbrace$.
Every node hosts the corresponding dataset, thus, $N$ geo-distributed datasets
are available as depicted by the following set 
$\mathcal{D} = \left\lbrace D_{1}, D_{2}, \ldots, D_{N} \right\rbrace$.
Datasets contain multivariate vectors reported by the IoT devices 
being connected with EC nodes.
$n_{i}$ hosts the reports of `its' IoT devices and formulates
a dataset $D_{i} = \{\mathbf{x}_{j}\}_{j=1}^{m_{j}}$ with $m_{j}$ real-valued 
contextual multidimensional data vectors.
Each data vector  
$\mathbf{x} = [x_{1}, x_{2}, \ldots, x_{d}]^{\top} \in \mathbb{R}^{d}$
involves the data related to $d$ dimensions. For instance, IoT devices may monitor
a phenomenon and report sensory data related to it (e.g., they could monitor a fire event and report data for temperature, humidity, etc). 
Any processing activity in $n_{i}$ is performed upon $D_{i}$ and targets to the provision of analytics or knowledge.
For instance, requests may demand for 
a regression analysis, classification, the estimation of multivariate and/or uni-variate histograms per attribute, non-linear statistical dependencies between input attributes and an application-defined output attribute, clustering of the contextual vectors,
etc. 
$D_{i}$s are also the basis for delivering the discussed synopses.
Let us denote a statistical synopsis by 
$\mathbf{s}$. 
$\mathbf{s}$ is depicted by $l$-dimensional vectors, i.e., 
$\mathbf{s} = [s_{1}, s_{2}, \ldots, s_{l}]^{\top} \subset \mathbb{R}^{l}$.
As mentioned above $\mathbf{s}_{i}$ delivered by $n_{i}$ is the summarization of 
$D_{i}$ upon the data reported by the corresponding IoT devices. 
Obviously, there are $N$ synopses $\mathbf{s}_{1}, \ldots, \mathbf{s}_{N}$ that have to be distributed in the ecosystem of EC nodes.

Let us focus on the behaviour of a specific EC node $n_{i}$. A similar approach dictates 
the behaviour of all the available nodes. 
Initially, $n_{i}$ is responsible to calculate 
locally the corresponding synopsis $\mathbf{s}_{i}$ upon $D_{i}$.
This happens when the received data change the statistics of the 
underlying dataset (e.g., concept drift). 
Afterwards, $n_{i}$ tries to act in a cooperative manner and decide to 
exchange $\mathbf{s}_{i}$ regularly. 
The target is to inform peers about the changes in the statistics of its dataset, thus, to give them the opportunity to be aligned with new trends in $D_{i}$.
Obviously, $n_{i}$, before sending $\mathbf{s}_{i}$, should take into 
consideration the trade off between the 
communication overhead and the `freshness' of $\mathbf{s}_{i}$ delivered to peers.
$n_{i}$ can share up-to-date synopses every time a change (even the smallest one) 
in the underlying data is realized
at the expense of flooding the network with numerous messages. 
We have to keep in mind that the connection of the IoT and EC infrastructures involves 
numerous devices exchanging numerous messages to convey data, synopses, knowledge, etc.
Hence, the frequency of the delivery of messages plays a significant role in the 
performance of the network.
In any case, a frequent delivery of $\mathbf{s}_{i}$ will give the opportunity to peers 
to enjoy fresh information
increasing the performance of decision making.
An intermediate solution is to postpone the delivery of $\mathbf{s}_{i}$
and reduce the sharing rate 
expecting less network overhead in light of `obsolete' synopses.
The delay in delivering $\mathbf{s}_{i}$ can be dictated 
by the limited updates in $\mathbf{s}_{i}$ as the result of retrieving 
data that cannot 
significantly change the statistics of $D_{i}$.
In this paper, we rely on the second approach and propose a model that 
monitors $\mathbf{s}_{i}$ and detects where significant changes 
in the underlying data are present before it decides to deliver the updated $\mathbf{s}_{i}$.
The target is to optimally limit the messaging overhead.
Our rationale is to monitor the `magnitude' of the 
collected statistical synopsis before we decide a dissemination action.
In this approach, there are two main problems.
The first is related to if past observations are 
the appropriate basis for initiating 
the delivery of $\mathbf{s}_{i}$ while the second deals with the uncertainty in the adopted 
threshold that will `fire' the delivery action.
Thresholds are set into any decision making mechanism that tries to detect the appropriate time to 
initiate an action.
For the first problem, we proposed the use of an LSTM/RNN capable of learning the dependencies of data, thus it will be easy to retrieve their future estimates. 
For the second problem, we proposed the use of a T2FLS to handle the incorporated uncertainty, i.e., our T2FLS results the DoD upon 
past synopsis observations and its estimated values. The proposed T2FLS tries to bridge the `gap' between past observations and future trends of synopses.
In any case, EC nodes are forced to disseminate synopses 
at pre-defined intervals even if no delivery decision is the outcome from 
our model.
We have to notice that, to avoid bottlenecks in the network, we consider the pre-defined intervals to differ among the group of EC nodes.
This `simulates' a load balancing approach avoiding to have too many EC nodes disseminating their synopses at the same time.

Our LSTM/RNN and T2FLS are fed by the most recent $\mathbf{s}_{i}$.
The RRN retrieves future estimates of $\mathbf{s}_{i}$ that are also fed 
into the proposed T2FLS. To the bast of our knowledge, the proposed model
is one of the first attempts that combines a DML with an FL system to deliver a powerful
decision making mechanism. 
The LSTM/RNN undertakes the responsibility of 
learning the data and their dependencies through time and 
the T2FLS focuses on the management of uncertainty in decision making.
$n_{i}$ monitors significant changes in $\mathbf{s}_{i}$ 
as more contextual data are received from IoT devices.
Based on the local monitoring activity, implicitly, we incorporate 
into the network edge the necessary `randomness' in the 
conclusion of the final decision, thus, potentially avoiding network flooding.
The discussed `randomness' is enhanced by different 
data arriving to the 
available nodes and their autonomous decision making. 
Such `randomness' can assist in limiting the possibility of deciding the delivery of synopses at the same time, thus, we can limit the possibility of overloading the network. 
We consider that at $t$ (a discrete time instance) a new $\mathbf{x}$
arrives in $n_{i}$.
Afterwards, the corresponding synopsis
$\mathbf{s}^{t-1}_{i}$ should be updated to conclude the 
new $\mathbf{s}^{t}_{i}$.
Let $\mathbf{e}_{t}$ be the difference over 
the current, last sent synopsis $\mathbf{s}^{t-1}_{i}$ and the new, the updated one,
$\mathbf{s}^{t}_{i}$.
We name this error
as the \textit{update quantum}, i.e., the magnitude of the difference between $\mathbf{s}^{t-1}_{i}$ and $\mathbf{s}^{t}_{i}$.
$n_{i}$ calculates $\mathbf{e}_{t}$ at consecutive 
time steps and, in a simplistic way, can be concluded by adopting the 
\textit{sum of differences} between two consecutive synopsis
for every dimension.
In any case, we can rely on any desired synopses realization technique.
$\mathbf{e}_{t}$ may have a positive or a negative trend, i.e.,  
the new vector can increase or decrease the value of each dimension.
For easiness in our calculations, we take into consideration the
absolute value of any difference into the available dimensions.
EC nodes should delay the delivery of $\mathbf{s}^{t}_{i}$
until they see that a significant difference, i.e., 
a high \textit{magnitude} depicted by $\mathbf{e}_{t}$
is true.
At that time, it is necessary to have the peer nodes informed 
about the new status of the local dataset.
We define the \textit{update epoch} as the time between disseminating $\mathbf{s}^{t-1}_{i}$
and $\mathbf{s}^{t}_{i}$. 
The update epoch is realized at pre-defined intervals,
$T, 2T, 3T, \ldots$ ($T >0$).
In this description, we focus on a single interval, 
e.g., $[1, 2, \ldots, T]$ where EC nodes 
check the last $\mathbf{e}$ realizations and feed them into our LSTM/RNN and T2FLS to 
see if they excuse the initiation of the dissemination process.
For sure, the dissemination of $\mathbf{s}^{t}_{i}$ will be concluded 
at $T$ if no relevant decision is made by our scheme.
$n_{i}$ also `reasons' over the time series of update quanta 
$\left\lbrace \mathbf{e}_{t} \right\rbrace$ with $t=1, 2, \ldots, T$.
It `projects' the time series to the future through the adoption of 
our LSTM/RNN. 
Again, the projection of update quanta is fed into the T2FLS to generate 
the DoD upon the future estimations of $\mathbf{e}$.

\section{Uncertainty Driven Proactive Synopses Dissemination}
\label{uncertainty}

\subsection{Estimating Future Trends of Synopses}




We select to adopt an LSTM \cite{goodfellow}, i.e., a specific type of RNNs to capture synopses trends for each dataset.
Our LSTM tries to `understand' every synopsis realization based on previous realizations and efficiently learn their distribution.
Legacy neural networks cannot perform well in cases where we want to capture the trend of a time series.
RNNs and LSTMs are network with loops inside of them making data to persist. 
We have to notice that the LSTM delivers $DoD_{f}$ for each synopsis realization.
In our model, we adopt an LSTM for the following reasons:
(i) we want to give the opportunity to the proposed model to learn over large sequences of data ($T >> 1$) and not only over recent data. Typical RNNs suffer from short-term memory and may leave significant information from the beginning of the sequence making difficult the transfer of information from early steps to the later ones;
(ii) typical RNNs also suffer from the \textit{vanishing gradient problem}, i.e., when a gradient becomes very low during back propagation, the network stops to learn;
(iii) LSTMs perform better the processing of data compared to other architectures as they incorporate multiple `gates' adopted to regulate the flow of the information. Hence, they can learn better than other models upon time series.

Every LSTM cell in the architecture of the network has an internal recurrence (i.e., a self-loop) in addition to the external recurrence of typical RNNs.
It also has more parameters than an RNN 
and the aforementioned gates to control the flow of data.
The self-loop weight is controller by the so-called forget gate, i.e.,
$g_{f}^{t} = \sigma \left( b^{f} + \sum_{j} U_{j}^{f} \mathbf{e}_{j}^{t} + \sum_{j} Z_{j}^{f} h_{j}^{t-1}\right)$   
where 
$\sigma$ is the standard deviation of the unit, $b^{f}$ represents the bias of the unit,
$U^{f}$ represents the input weights,
$\mathbf{e}$ is the vector of inputs
(we can get as many inputs as we want out of $W$ recordings),
$Z^{f}$ represents the weights of the forget gate and $h^{t-1}$ represents the current hidden layer vector.
The internal state of an LSTM cell is updated as follows:
$s^{t} = g_{f}^{t} s^{t-1} + g_{in}^{t} \sigma \left( b + \sum_{j} U_{j} \mathbf{e}_{j}^{t} + \sum_{j} Z_{j} h_{j}^{t-1}\right)$.
Now, 
$b$, $U$ and $Z$ represent the bias,
input weights and recurrent weights of the cell
and $g_{in}$ depicts the external input gate.
We perform similar calculations for the external input $g_{in}$ and the output gates
$g_{out}$. 
The following equations hold true:
\begin{equation}
g_{in}^{t} = \sigma \left( b^{in} + \sum_{j} U_{j}^{in} \mathbf{e}_{j}^{t} + \sum_{j} Z_{j}^{in} h_{j}^{t-1}\right)
\end{equation}

\begin{equation}
g_{out}^{t} = \sigma \left( b^{out} + \sum_{j} U_{j}^{out} \mathbf{e}_{j}^{t} + \sum_{j} Z_{j}^{out} h_{j}^{t-1}\right)
\end{equation}

The output of the cell is calculated as follows:

\begin{equation}
h^{t} = tanh \left( s^{t} \right) g_{out}^{t}
\end{equation}

We adopt a multiple input, single output LSTM.
In our case, we consider that the number of inputs/outputs are the three most recent synopsis error observations,
i.e., $\mathbf{e}_{t-2}, \mathbf{e}_{t-1}, \mathbf{e}_{t}$ for inputs and 
$\mathbf{e}_{t+1}, \mathbf{e}_{t+2}, \mathbf{e}_{t+3}$.
It should be noticed that our LSTM is trained upon real datasets by calculating 
the synopses of the reports as we reveal in our experimental evaluation section.
Past observations $\mathbf{e}_{t-2}, \mathbf{e}_{t-1}, \mathbf{e}_{t}$ are fed into the proposed T2FLS to retrieve the $DoD_{p}$ as well as future estimations  
$\mathbf{e}_{t+1}, \mathbf{e}_{t+2}, \mathbf{e}_{t+3}$ are adopted by our T2FLS
to retrieve the $DoD_{f}$.
Hence, our decision making model 
delivers the appropriate outcomes based on both 
approaches upon the statistical information of the 
local synopses.

\subsection{The Uncertainty driven Model}
For describing the proposed T2FLS, we borrow the notation of our previous efforts (in other domains)
presented in \cite{kolomvatsos2017}, \cite{kolomvatsos2016}.
T2FLS is adopted locally at every node
at $t$ 
by fusing 
the past $\mathbf{e}_{t}$ observations 
and future $\mathbf{e}_{t}$ realizations. 
$\mathbf{e}_{t}$
is adopted as the indication
whether the current update quanta 
significantly 
deviate from 
their past and future short-term trends. 
The envisioned fusion of update quanta is 
achieved through a finite set of
\textit{Fuzzy Rules} (FRs).
FRs incorporate past quanta or future estimations 
(two different processes) to reflect 
the $DoD$. Actually, we `fire' in two consecutive iterations 
the T2FLS for 
the last three (3) quanta realizations, 
i.e., $\mathbf{e}_{t-2}, \mathbf{e}_{t-1}, \mathbf{e}_{t}$ and the future three (3) 
quanta estimations, i.e., $\mathbf{e}_{t+1}, \mathbf{e}_{t+2}, \mathbf{e}_{t+3}$.
Our T2FLS, 
defines 
the fuzzy 
knowledge base 
for every $n_{i}$, 
e.g., a set of FRs like: 
`\textit{when the past/future quanta exhibit a significant 
difference from the last synopsis delivery,  
the $DoD$ for initiating the delivery of the new synopsis might 
be also high}'. 
We rely on Type-2 FL sets as 
the `typical' Type-1 fuzzy sets 
and the FRs defined upon them involve uncertainty due to
partial knowledge in representing 
the output of the inference \cite{ref45}.
The limitation 
of a Type-1 FL system is 
on handling uncertainty 
in representing knowledge 
through FRs \cite{ref28}, \cite{ref45}.
In such cases, uncertainty 
is observed 
not only in the environment, 
e.g., we classify 
the $DoD$ as `low' or `high', 
but also on the 
description of the term, e.g., `low'/`high', 
itself. 
In a T2FLS,  
membership functions  
are themselves `fuzzy',
which leads 
to the definition of FRs 
incorporating such uncertainty \cite{ref45}. 

FRs refer to a 
non-linear mapping
between three inputs:
(i) when focusing on the past quanta, we take as the following
as inputs into the T2FLS: 
$\mathbf{e}_{t-2}, \mathbf{e}_{t-1}, \mathbf{e}_{t}$;
(ii) when focusing on the future quanta, we take the following
asinputs into the T2FLS: 
$\mathbf{e}_{t+1}, \mathbf{e}_{t+2}, \mathbf{e}_{t+3}$.
The outputs are $DoD_{p}$ \& $DoD_{f}$, respectively.
The antecedent part of FRs 
is a (fuzzy) conjunction of inputs and 
the consequent part of the FRs 
is the $DoD$ indicating 
the belief that an event \textit{actually} occurs.
The proposed FRs have 
the following structure:
\noindent
\textbf{IF} $\mathbf{e}_{t-2}$ is $A_{1k}$ \textbf{AND} $e_{\mathbf{e}_{t-1}}$ is $A_{2k}$ \textbf{AND} $\mathbf{e}_{t}$ is $A_{3k}$ \\
\noindent 
\textbf{THEN} $DoD_{p}$ is $B_{k}$,
\noindent

\noindent
\textbf{IF} $\mathbf{e}_{t+1}$ is $A_{1k}$ \textbf{AND} $e_{\mathbf{e}_{t+2}}$ is $A_{2k}$ \textbf{AND} $\mathbf{e}_{t+3}$ is $A_{3k}$ \\
\noindent 
\textbf{THEN} $DoD_{f}$ is $B_{k}$,
\noindent

where $A_{1k}, A_{2k}, A_{3k}$ and $B_{k}$ are 
membership functions for the $k$-th FR 
mapping $\mathbf{e}_{i}, \mathbf{e}_{j}, \mathbf{e}_{k}$ and $DoD_{v}$,
$i \in \left\lbrace t-2, t+1 \right\rbrace$,
$j \in \left\lbrace t-1, t+2 \right\rbrace$,
$k \in \left\lbrace t, t+3 \right\rbrace$ and
$v \in \left\lbrace p, f \right\rbrace$.
For FL sets, we characterize their 
values through the terms: \textit{low}, \textit{medium}, and \textit{high}. 
The structure of  
FRs in the proposed T2FLS 
involve linguistic terms, e.g., \textit{high},  
represented by two membership functions, i.e.,
the 
\textit{lower} and the \textit{upper} bounds \cite{ref44}. 
For instance, the term `\textit{high}' 
whose membership for $x$ is a number $g(x)$, 
is represented by two membership functions defining the 
interval $[g_{L}(x), g_{U}(x)]$.
This interval 
corresponds to a lower and an upper membership function 
$g_{L}$ and $g_{U}$, respectively 
(e.g., the membership of $x=0.25$ can be in the interval $[0.05, 0.2]$). 
The interval areas $[g_{L}(x_{j}), g_{U}(x_{j})]$ for each  $x_{j}$ 
reflect the uncertainty in defining the term, e.g., `\textit{high}',  
useful to determine the exact membership function 
for each term. 
Obviously, if $g_{L}(x) = g_{U}(x), \forall x$, 
we obtain a FR in a Type-1 FL system. 
The interested reader could refer to \cite{ref44}
for information on reasoning under 
Type-2 FIRs.
We have to notice that FRs and membership functions for the proposed T2FLS are defined
by experts. 

\subsection{Synopses Update and Delivery}
As mentioned, in an iterative manner, our T2FLS is fed by 
past realizations and future estimations of the 
update quanta calculated upon the available datasets. 
The outcomes are depicted by $DoD_{p}$ and $DoD_{f}$.
We have to combine these two results into the final 
$DoD$ that exhibits the potential of initiating the delivery 
of the current synopsis.
In other words, we have to combine our view on the past with our estimations 
of the future before we decide to distribute the updated synopsis 
to peer nodes. 
For the aggregation process, we strategically select to 
rely on a simple methodology that will derive the final outcome in real time.
We propose the use of the geometric mean
\cite{mesiar} as the function for integrating the two aforementioned views on the
updated synopses.
The following equation holds true:
\begin{equation}
G(DoD_{p}, DoD_{f}) = \left( \prod_{i=1}^{2} DoD_{i} \right)^{1/2}
\end{equation}
with $i \in \left\lbrace p, f \right\rbrace$.
The rationale behind the adoption of the geometric mean is
that it is not affected by extreme values (high or low) 
and deals with all the inputs.
Moreover, the multiplicative approach 
supported by the geometric mean makes our model to be 
`strict' approach. FOr instance, when one the two DoD values is zero, the final outcome is zero as well. 
This way, we want to be sure that there is 
`critical' amount of magnitude in synopses before 
they a re distributed in the network.
The final decision depends on a threshold $\theta$.
When $G>theta$, we initiate the dissemination action.
$\theta$ is a pre-defined threshold that `dictates' when 
an EC node should pursue the exchange of a synopsis.

\section{Experimental Setup and Evaluation}
\label{evaluation}

\subsection{Setup and Performance Metrics}
We report on the performance of our \textit{Uncertainty Driven
Synopses Dissemination Model} (UDSDM) and compare it 
with other baseline models and schemes proposed in the relevant literature.
Initially, we focus on the  
percentage of $T$ that our model spends till the final decision.
The $\phi$ metric is defined as follows:
$\phi = \frac{1}{E} \sum \left\lbrace \frac{t^{*}}{T} \right\rbrace_{i=1}^{E}$
where $t^{*}$ is the time when the dissemination actions is decided,
$E$ is the number of  
experiments and 
$i$ depicts the index of every experiment. 
When $\phi \to 1$ means that the 
proposed model spends the entire interval $T$ 
to conclude a final decision.
When $\phi \to 0$, our model manages to 
conclude immediately the dissemination action.
Additionally, we define 
the metric $\delta$ i.e., 
$\delta = \frac{1}{E} \sum \left\lbrace |\mathbf{s}^{t^{*}} - \mathbf{s}| \right\rbrace_{i=1}^{E}$.
$\delta$ represents the average magnitude 
of the difference between the current and the 
new synopses.
Through $\delta$, we want to depict the 
ability of the proposed model to `react' even in 
limited changes in the updated synopses 
(we target a $\delta \to 0$).
The magnitude is calculated 
at $t^{*}$.
The ability of the proposed system to avoid the overloading of the network and limiting the required number of messages is exposed by  
$\psi$. 
The following equation holds true:
$\psi = \frac{T}{|t^{*}|_{t^{*} \in [1,T]}}$ ($\psi \in [0,T]$)
where $|t^{*}|_{t^{*} \in [1,T]}$ represents the number of 
times that the model stops 
in the interval $[1,T]$.
When $\psi \to 1$ means that the proposed model 
stops frequently, thus, multiple messages conveying the
calculated synopses are transferred through the network.
When $\psi \to T$ means that our model does not stop frequently, thus, the calculated synopses are delivered close to the expiration of $T$.
For our experimentation, we adopt the dataset 
presented in Intel Berkeley Research Lab \cite{chu}. 
It contains measurements from 54 sensors deployed 
in a lab. We get the available measurements and simulate the provision 
of context vectors to calculate the synopses and the update quanta (they are realized
in the interval $[0, \infty]$)
in a sequential order. 
We also pursue a comparative assessment for the UDSDM with:
\textbf{(i)} a baseline model (BM) that 
disseminates synopses when any change is observed over 
the incoming data; 
\textbf{(ii)} the Prediction based Model (PM)
\cite{martin}: 
PM proceeds with the stopping decision only when 
the estimation of the future update quanta violates 
a threshold. For realizing the PM, we adopt the double exponential smoothing 
method \cite{vandeput}. The method applies a recursive model of an exponential filter twice before it results the final outcome. 
We perform simulations for $W = 10$ adopted to realize the double exponential 
smoothing scheme and $T \in \left\lbrace 100, 500, 1000 \right\rbrace$. 
In every experiment, we run the UDSDM and get numerical results related to the mean of the aforementioned metrics (we adopt $\theta \in \left\lbrace 0.60, 0.75 \right\rbrace$ for the UDSDM and the PM). 

\subsection{Performance Assessment}
\begin{figure}[h]
\centerline{\includegraphics[width=95mm,scale=1.0]{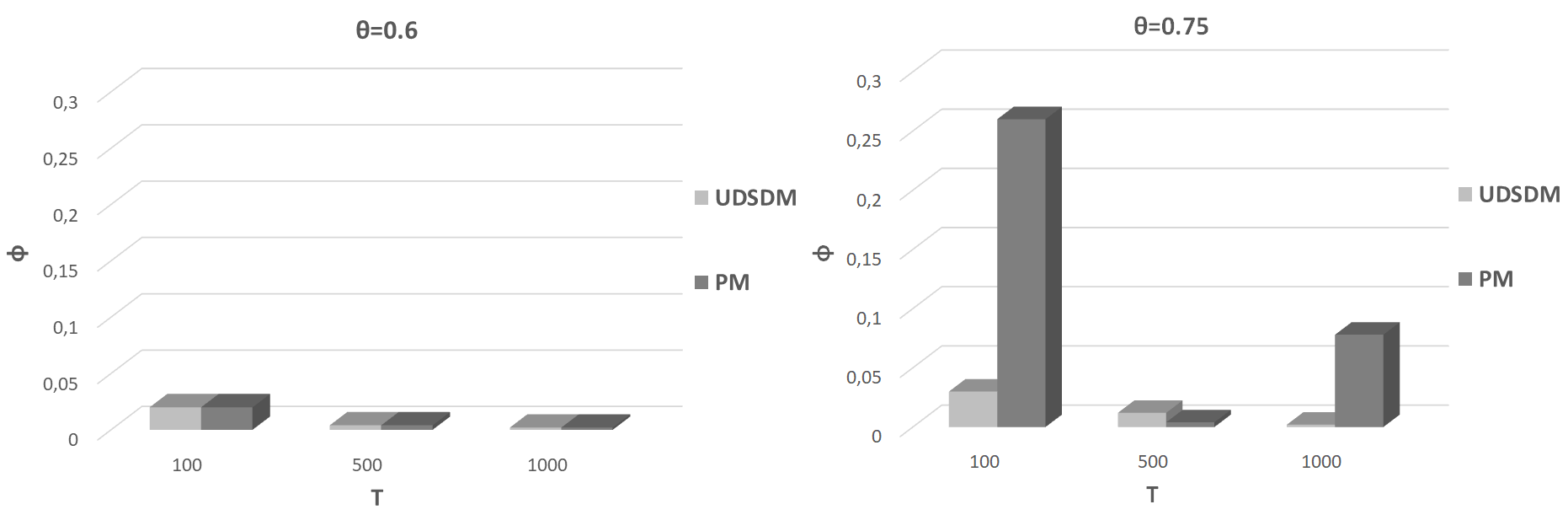}}
\caption{Comparative results for the $\phi$ metric.}
\label{fig1}
\end{figure}

In Fig. \ref{fig1}, we present our results for the $\phi$ metric. We observe that the adoption of a low $\theta$ (threshold for deciding the 
dissemination action) leads to an decreased time for the final decision. This means that the proposed
model manages to conclude immediately a fuzzy result upon $\theta$ that `fires' the dissemination action.
In addition, and increased $T$ leads to a decreased 
$\phi$. The higher the $T$ is, the lower the $\phi$ becomes. 
When $\theta$ and $T$ are high, the percentage of $T$ devoted to conclude the dissemination decision is very low. Compared to the PM, the UDSDM requires less time to conclude the delivery action (except when $\theta = 0.75$ \& $T = 500$) for the majority of the experimental scenarios.
Actually, the proposed system manages to deal with the final decision as soon as it detects that update quanta are aggregated over time even in small amounts. This can be realized in early monitoring rounds due to the dynamic nature of the incoming data.
Recall that we adopt a time series that consists of sensory data retrieved by a high number of devices that are, generally, characterized by their dynamic nature.

\begin{figure}[h]
\centerline{\includegraphics[width=95mm,scale=1.0]{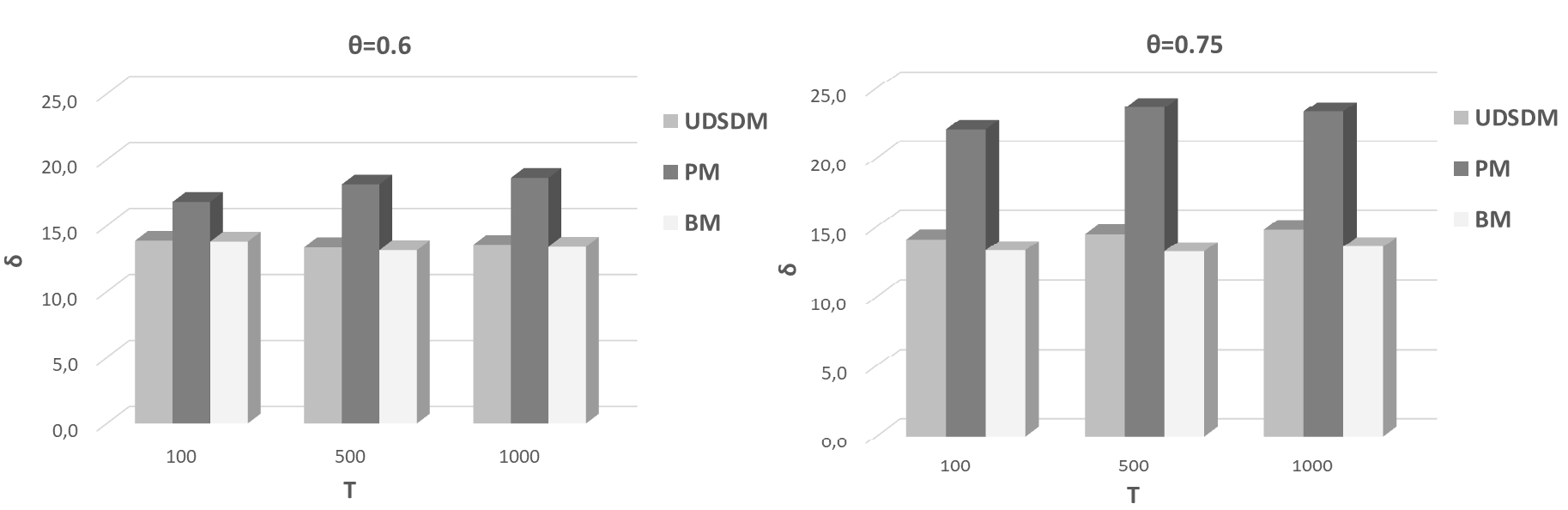}}
\caption{Comparative results for the $\delta$ metric.}
\label{fig2}
\end{figure}

Fig. \ref{fig2} presents our results related to the $\delta$ metric,
i.e., the update quanta at the time when the dissemination action is decided.
We observe that the UDSDM requires a lower magnitude than the PM and higher or equal than the BM before it concludes the dissemination action. When 
$\theta = 0.6$, there is `stability' of the 
required $\delta$ before the dissemination action.
When $\theta = 0.75$, $\delta$ increases together 
with $T$. The PM requires a higher synopses magnitude to be collected compared to the remaining two models.
These result present the `attitude' of the proposed model to wait and aggregate update quanta in order to alleviate the network from an increased number of messages. However, 
our model does not wait till the expiration of $T$ to report fresh synopses to the network.
We can easily observe that the proposed model relies 
in the middle between the BM and the PM (with an `attitude' to be close to the BM).    

\begin{figure}[h]
\centerline{\includegraphics[width=95mm,scale=1.0]{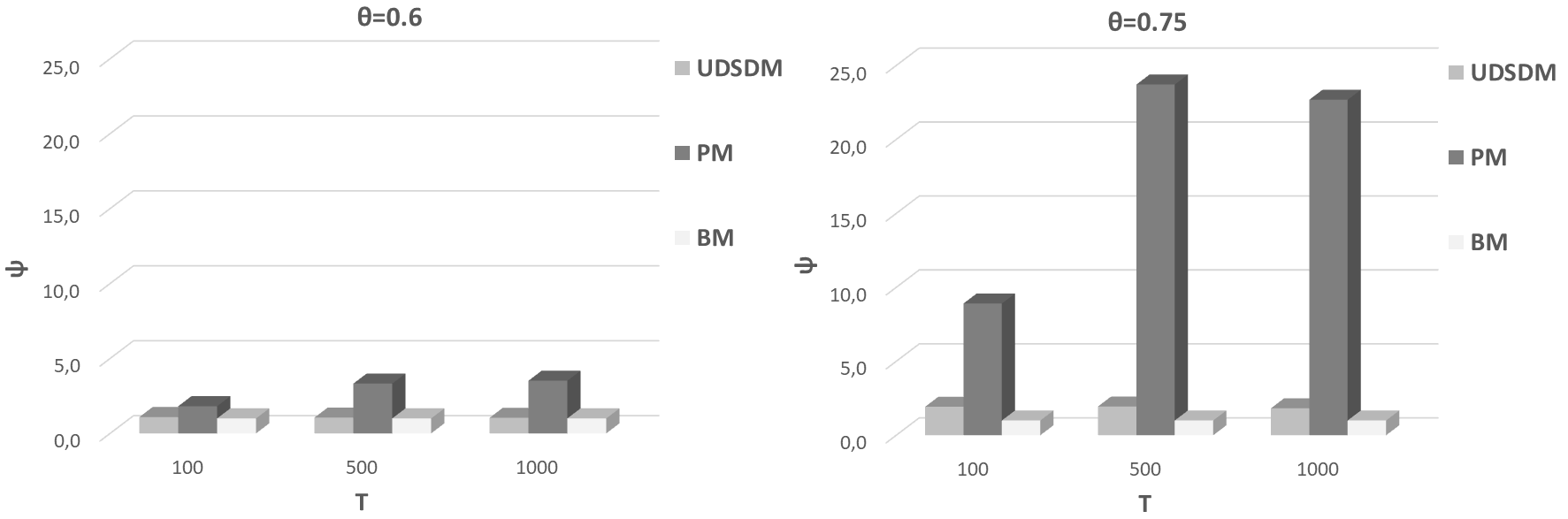}}
\caption{Comparative results for the $\psi$ metric.}
\label{fig3}
\end{figure}

In Fig. \ref{fig3}, we present our results related to the 
$\psi$ metric. 
We confirm our observations obtained by the two above discussed metrics, 
i.e., the UDSDM relies in the middle between the BM and the PM.
Our model is mainly affected by the rationale to 
distribute fresh synopses in the burden of the 
number of messages circulated in the network. 
However, it manages to deliver less messages than the BM. We observe a stability in the obtained outcomes exhibiting the capability of the UDSDM to detect changes in synopses
quanta and fire the delivery action. The PM requires the less frequency of the delivery, however, in the burden of the freshness of the distributed synopses.

\section{Conclusions}
\label{conclusions}
Data management at the edge of the network is a significant research subject 
due to the reduced latency that end users can enjoy if processing is performed at the EC
infrastructure.
Numerous IoT devices report data towards the Cloud datacenter, thus, 
advanced data management applications should be provided at the EC 
as the intermediate point where processing can take place.
A number of EC nodes may undertake the responsibility of hosting datasets and 
processing activities.
Nodes should act in a collaborative manner to increase their performance.
For instance, EC nodes may exchange processing tasks or data 
to conclude the desired outcomes as soon as possible.
In this paper, we enhance the collaborative aspect of the EC infrastructure and propose
a novel model for exchanging data synopses at the edge of network. 
The target is to have all EC nodes informed about the data present at their peers, thus, 
to take optimal decisions related to the management of the requested processing activities.
We present a deep learning model and an uncertainty driven 
scheme to reason over the appropriate time to exchange data synopses. 
The deep learning model manages to learn the distribution of the concluded synopsis as the basis
for retrieving future estimations.
The uncertainty driven scheme deals with a set of rules applied upon past synopses observations and future estimates.
This way, we combine two completely different technologies to 
realize an efficient system for the management of data synopses at EC.
Our aim is to provide a decision making methodology that minimizes the number of messages circulated in the network, however, without jeopardizing 
the freshness of the exchanged statistical information.
We discuss our model adopting the principles of the FL
and present the relevant formulations.
EC nodes monitor their data and decide when it is
the right time to deliver the current data synopsis. 
Our experimental evaluation shows that the proposed scheme can efficiently assist 
in the envisioned goals being evidenced by 
numerical results. 
In the first place of our future research plans, it is to incorporate a 
rewarding mechanism for every `correct' decision and present a system that learns on how to learn. 
Additionally, we want to involve more parameters in the decision making 
mechanism like a `snapshot' of the current status of every EC node.


\end{document}